\documentclass[11pt]{article}
\usepackage{geometry}                
\usepackage{graphicx}
\usepackage{amssymb}
\usepackage{epstopdf}
\usepackage{caption}
\usepackage{threeparttable}
\usepackage{array}

\usepackage[colorlinks = true,
            linkcolor = blue,
            urlcolor  = blue,
            citecolor = blue,
            anchorcolor = blue,
            hypertexnames=false] 
            {hyperref}

\def\authorlist#1#2{
    \vskip 0.4in
\begin{center}\begin{large} {\bf  #1 } \end{large}
    \vskip 0.2in
              #2
     \vskip 0.2in
   \end{center}
}

\begin{document}

\title{\textbf{Snowmass 2021 \\ Underground Facilities \& Infrastructure \\ Frontier Report}}

\maketitle

\authorlist{L.~Baudis, J.~Hall, K.T.~Lesko, J.L.~Orrell}{}

\tableofcontents




%
%
\section*{Executive Summary}
\label{sec:uf-executive}

From the properties and nature of the neutrino to the direct measurement of galactic halo dark matter, research performed at underground facilities tackles a collection of precision knowledge of the constituents of the Standard Model of particle physics to broadly open-ended search well Beyond the Standard Model (BSM). The chapters and reports from other Frontiers layout the science case for these and other avenues of particle physics research. Here we identify the needs and requirements for underground facilities continue to support the breadth of particle physics research planned in the coming 10--15 years.

Our overall evaluation begins with assessment starting with the Underground Laboratory Capabilities report from Snowmass 2013~\cite{10.48550/arxiv.1401.6115} and those 2014 P5 report~\cite{osti_1320565} recommendations relating to science performed in underground facilities. The Snowmass report conclusions included locating LBNF/DUNE at the Sanford Underground Research Facility, pursue continued leadership in dark matter and neutrinoless double-beta decay, coordinate with overseas and domestic underground facilities to ensure a position of leadership, and maintain a domestic underground facility to support current and future generations of largest dark matter and neutrinoless double-beta decay experiments. The 2014 P5 report recommended proceeding with LBNF/DUNE in the U.S., launch the Generation 2 dark matter direct detection program, and then plan and initiate a follow-on Generation 3 dark matter direct detection program.

The underground facilities and infrastructure Topical groups assessed the requirements in the areas of neutrino, cosmic frontier, supporting capabilities, synergistic, non-high energy physics research also performed in underground facilities, and an assessment of current status of underground facilities world-wide. While numerous observations resulted, several key observations set the framework for planning for the next decade of science at underground facilities:
\begin{itemize}
    \item The LBNF/DUNE program of neutrino science will be the flagship U.S. high energy physics research effort for the coming decade and beyond. This program hinges on success at the Sanford Underground Research Facility (SURF).
    \item The long-term underground infrastructure investment at SURF creates opportunity for complementary and synergistic research, both within the high energy physics community and beyond.
    \item The current programs of dark matter direct detection will culminate in the late 2020s and planning for next generation dark matter research should begin in this Snowmass period to ensure the underground facilities and infrastructure are available and ready.
    \item While much of the required expertise and infrastructure for future underground high energy physics research is well developed, current underground facilities are fully subscribed with existing or planned experiments. New space (excavation) is likely required to support the programs of research for the coming decade and beyond.
\end{itemize}

From the above assessments, the primary conclusions of the Underground Facilities and Infrastructure evaluation are:
\begin{itemize}
    \item[]\textbf{UF \#1:} Leverage the Long Baseline Neutrino Facility excavation enterprise to increase underground space at Sanford Underground Research Facility in a timely and cost-effective way to permit siting of next-generation underground high energy physics research experiments.
    \item[]\textbf{UF \#2:} Designate the Sanford Underground Research Facility as a U.S. Department of Energy User Facility.
    \item[]\textbf{UF \#3:} Provide full support for the underground facilities hosting the Long Baseline Neutrino Facility (LBNF) and the Deep Underground Neutrino Experiment (DUNE).
    \item[]\textbf{UF \#4:} Following the 2014 P5 Recommendation 20, R\&D and decision making for a third-generation direct detection dark matter program should commence immediately to enable a construction start in the late 2020s.
    \item[]\textbf{UF \#5:} To ensure a robust collection of scientific programs in underground facilities, support the enabling capabilities, technique development, and expertise required for underground experiments.
\end{itemize}
Within the full report below of from the Underground Facilities and Infrastructure Frontier, each of these conclusions is explained in context and presented in greater detail.

%
%
\section{Introduction}
\label{sec:uf-introduction}

The decade since Snowmass 2013 has seen extraordinary progress of high energy physics research performed--or planned for--at underground facilities. Drs. T. Kajita and A.B. McDonald were awarded the 2015 Nobel Prize in Physics for the discovery of neutrino oscillation~\cite{NobelPrize:2015-Physics}, which show that neutrinos have mass. The U.S. has embarked on the development of the world-class LBNF/DUNE science program to investigate neutrino properties. The Generation 2 dark matter program is advancing to full data collection in the coming $\sim$5~years, a Dark Matter New Initiatives program has begun, and the U.S. dark matter community is looking toward a Generation 3 program of large-scale dark matter direct detection searches. The Sanford Underground Research Facility has become a focal point for U.S. underground facilities and infrastructure investment. Here we briefly review the status as was seen by those who participated in the 2013 Snowmass process as well as the outcome from the 2014 P5 program of recommendations.

\subsection{Taking stock of Snowmass 2013}

The Underground Laboratory Capabilities report from Snowmass 2013~\cite{10.48550/arxiv.1401.6115} outlines plans, needs, and concerns contemporary at that time. Here we briefly take stock of what has been accomplished in view of the four main conclusions provided by the Underground Laboratory Capabilities report. 

\begin{enumerate}
    \item Locate LBNE underground to realize its full science potential. This step would also provide a natural base for additional domestic underground capabilities at SURF in the future.
    \item The U.S. has leading roles in many of the future dark matter, neutrinoless double beta decay and neutrino experiments.
    \item More coordination and planning of underground facilities (overseas and domestic) is required to maintain this leading role, including use of existing U.S. infrastructure and closer coordination with SNOLAB as the deepest North American Lab.
    \item Maintaining an underground facility that can be expanded to house the largest dark matter and neutrinoless double beta decay experiments would guarantee the ability of the U.S. to continue its strong role in the worldwide program of underground physics.
\end{enumerate}

Since Snowmass 2013, LBNE has been split into the Long Baseline Neutrino Facility (LBNF) which provides the neutrino beam and facility complex and the Deep Underground Neutrino Experiment (DUNE) which is the suite of detector modules used to study neutrino properties~\cite{https://doi.org/10.48550/arxiv.1601.05471,https://doi.org/10.48550/arxiv.1512.06148,https://doi.org/10.48550/arxiv.1601.05823,https://doi.org/10.48550/arxiv.1601.02984}. Effectively, the first item is being fulfilled through the construction of LBNF and DUNE.

The second conclusion item is written as an observation, but largely remains true today. The DOE Office of High Energy Physics has directly supported the construction of two second-generation (G2) direct detection dark matter experiments located in underground facilities. The LZ experiment~\cite{https://doi.org/10.48550/arxiv.1703.09144}, located at the Sanford Underground Research Facility (SURF), has recently reported initial and world leading results~\cite{https://doi.org/10.48550/arxiv.2207.03764}. The SuperCDMS experiment~\cite{PhysRevD.95.082002}, located at SNOLAB, is under construction with data collection expected to commence in 2023. The DOE Office of Nuclear Physics has recently announced the intent to develop three world-leading, ton-scale neutrinoless double-beta decay experiments (i.e., CUPID~\cite{https://doi.org/10.48550/arxiv.1907.09376}, LEGEND-1000~\cite{https://doi.org/10.48550/arxiv.2107.11462}, and nEXO~\cite{https://doi.org/10.48550/arxiv.1805.11142}). As described above, LBNF/DUNE will become a world-leading effort in neutrino physics experiments.

Over the last decade, the U.S. has maintained its leadership role in the targeted underground research areas of ``dark matter, neutrinoless double beta decay and neutrino experiments''. Since the writing of the 2013 Snowmass reports SURF and SNOLAB have developed into world-recognized peers in underground science. Both of these underground facilities, as observed through the 2021--2022 Snowmass process, are operating ``at capacity'' with current and near-term planned experiments.

The fourth item is only partially complete. As just previously stated, SURF is operating ``at capacity'' with current and near-term planned experiments. However, expansion of SURF underground space would make the theme of the fourth item feasible on an appropriate time-scale (mid- to late-2020s) to fulfill the fourth conclusion.

\subsection{Progress since the 2014 P5 report}

The 2014 P5 report~\cite{osti_1320565} contained three recommendations relevant to assessing underground facilities and infrastructure:

\begin{itemize}
    \item[]\textbf{Recommendation 13:}~Form a new international collaboration to design and execute a highly capable Long-Baseline Neutrino Facility (LBNF) hosted by the U.S. To proceed, a project plan and identified resources must exist to meet the minimum requirements in the text. LBNF is the highest-priority large project in its timeframe. 
    \item[]\textbf{Recommendation 19:}~Proceed immediately with a broad second-generation (G2) dark matter direct detection program with capabilities described in the text. Invest in this program at a level significantly above that called for in the 2012 joint agency announcement of opportunity.
    \item[]\textbf{Recommendation 20:}~Support one or more third-generation (G3) direct detection experiments, guided by the results of the preceding searches. Seek a globally complementary program and increased international partnership in G3 experiments.
\end{itemize}

As described above, the LBNF/DUNE and the G2 dark matter programs are underway and largely fulfill \textbf{Recommendation 13} and \textbf{Recommendation 19}. However, \textbf{Recommendation 20} is at best in a nascent state.

%
%
\section{Key Observations from Underground Facilities \& Infrastructure Topical Reports}
\label{sec:uf-topical-summary}

In additional to the assessment described above of the progress since Snowmass 2013 and the 2014 P5 report, the 2021--2022 Snowmass process has freshly assessed the status needs for underground facilities and infrastructure to support the science goals of the next decade and beyond. In this section we briefly summarize the key observations stemming from the assessments performed by the Topical groups within the Underground Facilities and Infrastructure Frontier.

\subsection{Key Observations -- Underground Facilities for Neutrinos}

The LBNF/DUNE program is a key, flagship high energy physics research program relying on underground facilities and infrastructure. Notably the Sanford Underground Research Facility (SURF), where potions of LBNF/DUNE are located, is where all major U.S. investments in underground facilities and infrastructure have been made over the last 10 years.

The underground cavern space provided by LBNF will soon complete. The timeline for Phase II of DUNE is not yet decided, but likely initiates in the 2030 or later time frame. There are several potential scenarios for underground space utilization in advance of the completion of DUNE via Phase II construction.

Proposals for future large-scale detectors targeting measurement of natural neutrino sources (\textit{e.g.}, supernova, solar, geoneutrinos, \ldots) which, if pursued, will require large underground spaces which are not currently available in the suite of existing underground labs.

Future neutrinoless double-beta decay experiments, supported by nuclear physics programs, are expected to require underground facility space and infrastructure in the coming decade and beyond. It remains an open question whether deeper experimental locations are required for future neutrinoless double-beta decay experiments.

\textit{The full Underground Facilities for Neutrinos (UF1) topical report is available~\cite{https://doi.org/10.48550/arxiv.2209.07622}.}

\subsection{Key Observations -- Underground Facilities for the Cosmic Frontier}

Multiple new underground dark matter experiments are expected and being planned (at both large and small scales).  At the same time, underground facilities are largely subscribed by existing projects, with only very limited space available in the coming years.  There is, then, a clear need for additional underground space, tailored to the needs of dark matter experiments.

This underground space should specifically include both large spaces for large experiments (liquid noble) and small spaces for smaller experiments (cryogenic bolometer, ‘other’).  The assembly of large liquid noble experiments will occur largely in the underground environment, meaning they require large underground radon-free clean rooms.  Given the volume of gas/cryogen, such experiments also require large underground areas for staging (e.g., gas storage) and experiment utilities (e.g. pumps, distillation).

These new suitable spaces must be available by the late 2020's to meet the demand.  This demand may be met in North America by proposed new excavations at SURF or SNOLAB.

\textit{The full Underground Facilities for the Cosmic Frontier (UF2) topical report is available~\cite{UF2-forthcoming}.}

\subsection{Key Observations -- Supporting Capabilities for Underground Science}

Future, larger experiments will increasingly require underground assembly with stricter radioactivity requirements.  There will need to be larger, cleaner cleanrooms, often with better radon-reduction systems and increased monitoring capabilities for ambient contaminants.   
Methods to assay dust deposition and radon-daughter plate-out will need to be improved. There will be increased need for underground machine shops. Additionally, many labs are considering adding underground copper electroforming to mitigate against cosmogenic activation, as has been successfully demonstrated at SURF.

Most assay needs may be met by existing worldwide capabilities with organized cooperation between facilities and experiments. Improved assay sensitivity is needed for assays of bulk and surface radioactivity for some materials for some experiments, and would be highly beneficial for radon emanation.  

\textit{The full Supporting Capabilities for Underground Facilities (UF4) topical report is available~\cite{https://doi.org/10.48550/arxiv.2209.07588}.}

\subsection{Key Observations -- Synergistic Research in Underground Facilities}

A variety of research beyond high energy physics is performed in underground laboratories. These research activities from other disciplines include burgeoning R\&D in quantum information science, accelerator-based nuclear astrophysics, tests of fundamental symmetries, gravitational wave detection, and geology \& geophysics.

In most all cases, research groups from differing disciplines are able to work in shared underground facility space harmoniously and in some cases synergistically. Smaller-scale research efforts benefit substantially from the investment in underground facilities and infrastructure initially prepared for larger-scale science endeavors.

Evaluating the ``total research impact'' of shared-usage underground research facilities should take into account the full disciplinary breadth of research performed.

\textit{The full Synergistic Research in Underground Facilities (UF5) topical report is available~\cite{https://doi.org/10.48550/arxiv.2210.03145}.}

\subsection{Key Observations -- Underground Facilities \& Infrastructure Overview}

The 2022 Snowmass process documents an increasing demand for the special environments provided by deep underground laboratories to conduct high priority science.  Emergent applications have been identified beyond neutrinos and dark matter searches, expanding the demand across more HEP frontiers including quantum information science (QIS), quantum computing (QC), and atom interferometry. The 2014 P5 report established several high priorities for high energy physics including a domestic G3 direct detection dark matter effort and LBNF/DUNE.  The intervening years have witnessed a rapid growth in new technologies for a variety of HEP topics which are rapidly approaching prototype phases.  Some of these are a result of Dark Matter New Initiatives (DMNI) and Basic Research Needs (BRN) for High Energy Physics Detector Research \& Development efforts.

\subsubsection{Underground laboratories}

\begin{figure}[ht!]
    \centering
    \includegraphics[width=0.9\textwidth]{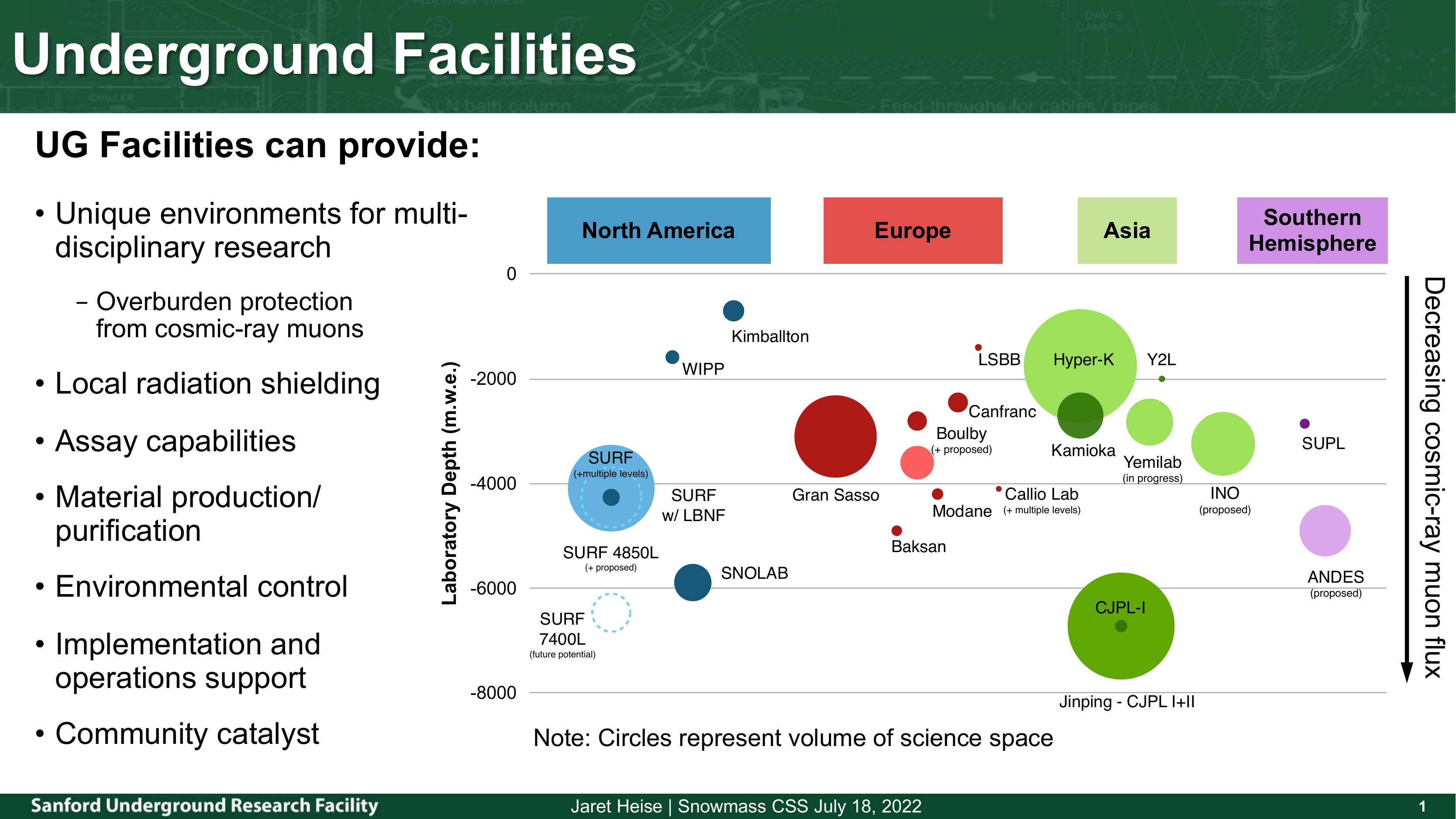}
    \caption{Overview of underground facilities located around the world, showing various key characteristics in depth and underground laboratory capacity. Slide from Jaret Heise (SURF).}.
    \label{fig:lab-depth}
\end{figure}

As part of Underground Facilities Frontier the major underground laboratories (see Fig.~\ref{fig:lab-depth}) were assessed for their current capacity to support science, current expansion efforts, and plans for future expansions.  These are summarized in Table~\ref{tab:ug-labs}.
The analysis identifies that most facilities have existing capacity to host small scale efforts including R\&D deployments and prototype experiments. Two facilities are expanding their facilities to accommodate long baseline neutrinos - LBNF/DUNE at SURF~\cite{https://doi.org/10.48550/arxiv.2203.08293} and Hyper-Kamiokande at Kamioka~\cite{https://doi.org/10.48550/arxiv.2203.13979}.  SNOLAB is reserving a significant cavity for neutrinoless double-beta decay and INFN-Gran Sasso is expanding their double-beta decay suite of experiments. The three Office of Nuclear Physics supported efforts, nEXO, LEGEND, and CUPID are in discussion or developing plans to utilize these facilities, lacking a domestic option for hosting these efforts with significant U.S. participation or leadership. The U.S. underground facility, SURF, does not have capacity to host a Generation 3 direct detection dark matter experiment.  Of the facilities surveyed by the Underground Facilities and Infrastructure group, LNGS and SNOLAB will have space opening following decommissioning of existing experiments. At the time of the Snowmass meeting in Seattle (July 2022), no Generation 3 experimental collaboration had assessed or selected an underground location.

\begin{table}
    \centering
    \small
    \begin{tabular}{ | p{0.08\linewidth} | p{0.07\linewidth} | >{\raggedright\arraybackslash}p{0.20\linewidth} | >{\raggedright\arraybackslash}p{0.1\linewidth} | >{\raggedright\arraybackslash}p{0.075\linewidth} | >{\raggedright\arraybackslash}p{0.11\linewidth} | >{\raggedright\arraybackslash}p{0.14\linewidth} | }
    \hline
    Facility & Depth (m.w.e.) & Current science program & Expansions underway & Capacity for small scale experiments & Capacity for G3 DM or tonne scale DBD & Proposals for future expansion \\
    \hline \hline
    Boulby & 2850 & CYGNUS, Low Background Assay, NEWS-G/Dark Sphere R\&D &  & modest & none & $\sim$30\,m diameter cavity for G3 DM. Potentially deeper location. \\ \hline
    Kamioka & 2700 & SuperK, T2K, KamLAND-Ze, CLIO, KAGRA, NEWAGE, CANDLES, Hyper-K & Hyper-K & modest & none & \\ \hline
    LNGS* & 3600--3800 & COBRA, COSINUS, CRESST, CUORE, CUPID-0, CUPID, DAMA, DARKSIDE, ERMES, GERDA, GINGER, LUNA, LVD, NEWS-DM, ERMES, GERDA, GINGER, LUNA, LVD, NEWS-DM, SABRE, VIP, XENON & & modest & Decom-missioning of Borexino, CTF, and DarkSide 50 ($\sim$20x20\,m$^2$) creates a large enough space for a G3 DM experiment & \\ \hline
    Sanford Underground Research Facility (SURF) & 4200 & LZ (DM), MJ-$^{180}$Ta, CASPAR, Low Background Assay, Geomicrobiology, Geoengineering, Education \& Outreach, LBNE/DUNE & LBNF and DUNE & limited & none & 100\,m lab module(s) \\ \hline
    SNOLAB & 6000 & SNO+(LS), PICO~40, CUTE, SENSEI, OSCURA, DAMIC, HALO, SuperCDMS SNOLAB, DEAP-3600 II, PICO-500, ECUME, NEWS-G, SNO+(Te), SBC &  & modest & Decom-missioning Cube hall projects creates a 18x15\,m space for G3 DM or tonne-scale DBD. & \\ \hline
    Yemilab & 2500 & LSC, AMoRE, COSINE, KNU, KIGAM, IsoDAR &  & limited & none & \\ \hline
    \end{tabular}
    \caption{Tabulation of key details of underground laboratories. A key focus is given to Generation 3 dark matter (G3 DM) and tonne-scale neutrinoless double-beta decay (tonne-scale DBD) as these classes of experiments have similar large space and infrastructure requirements. Depth is given in units of meters of water equivalent (m.w.e). * LNGS did not return the Snowmass UF questionnaire, however the September 2022 LNGS Science Committee report suggests availability of the Borexino and DarkSide spaces in Hall C. The assessment of available space for small scale experiments was approximately gauged by the metric: limited $\sim\leq$100\,m$^2$ and moderate between $\sim$100 and $\sim$500\,m$^2$.}
    \label{tab:ug-labs}
\end{table}

\subsubsection{Future underground laboratory capacity}

During the Snowmass process proposals were presented for expanding three laboratories to create adequate capacity for a G3 experiment. These are SURF(U.S.), SNOLAB(Canada), and Boulby(UK).  All three proposals would be adequate for supporting a G3 program.  The SURF proposal would provide space and capacity beyond those required by this one experiment.  The alternative off-shoring U.S. scientific efforts creates the potential for negative impacts on U.S. scientific leadership, technology development, and while increasing costs and adding challenges to the U.S. science. This was a recurring theme of discussions during Snowmass.  Pressure on the U.S. research budgets are amplified as a consequence of the significant travel costs and any international travel restrictions (e.g., COVID, etc.).

\begin{figure}[ht!]
    \centering
    \includegraphics[width=0.9\textwidth]{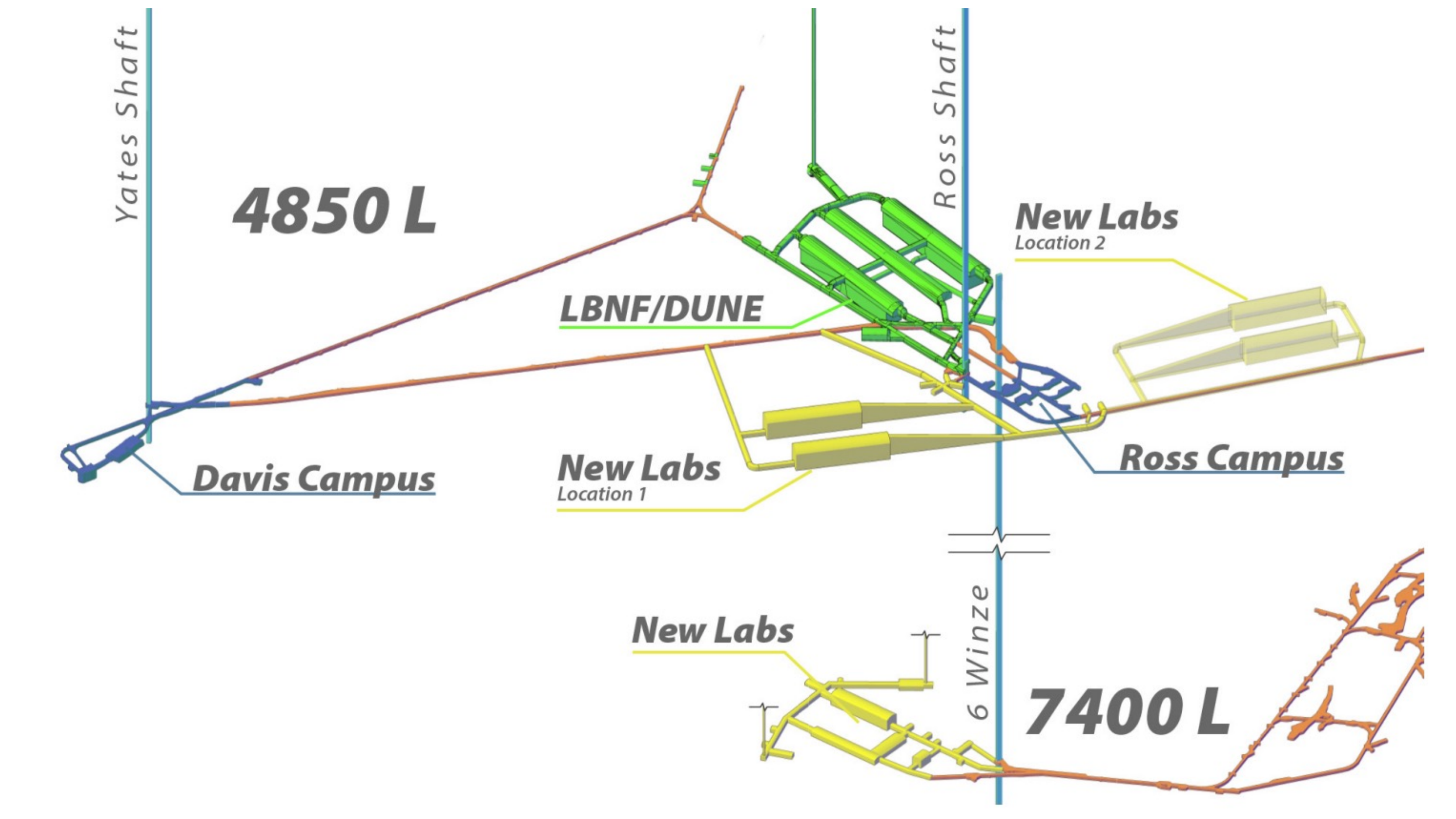}
    \caption{SURF expansion plan, with new labs labeled and shown in yellow. Graphic from Jaret Heise (SURF).}.
    \label{fig:SURF-expand}
\end{figure}

The SURF proposal is presented in the figure below.  Two locations for multiple laboratory modules have been developed near the Ross Campus, the site of current LBNF/DUNE excavations.  Excavation of these modules could utilize the significant infrastructure investments from the DUNE excavation to create these spaces.  Coordinating the lab modules with the DUNE contracts could result in very significant cost and schedule saving by reducing or eliminating mobilization and demobilization costs, O(\$15M).  A long term option for a deeper campus is also shown, though it was generally not considered as part of planning ``for the next Snowmass period''. Nevertheless, the deep lab option provides potentially opportunity for future growth should the scientific community's goal require increased shielding from cosmic ray secondaries.

The South Dakota Science and Technology Authority presented in their Snowmass whitepaper these conceptual plans and their intention to pursue State and Private funding for the excavation for one module. With appropriate coordination the 100\,m laboratory module would be ready for Beneficial Occupancy on a relevant timescale.  This would provide adequate space for a Generation 3 dark matter experiment, another large-scale experiment such as neutrinoless double-beta, or multiple smaller scale R\&D projects or pathfinder experiments.

%
%
\section{A U.S. strategy for underground facilities -- Conclusions}
\label{sec:uf-strategy}

The Underground Facilities and Infrastructure Frontier was charged to ``develop a integrated strategy for underground facilities in the coming decade and beyond''. A strategy requires a set of actionable steps. Thus, we frame the proposed strategy as a set of actionable conclusions, with clear steps for implementation, that will lead to outcomes substantially advancing and supporting the long-term vitality, capacity, and leadership potential of the U.S. high energy physics community to perform research requiring underground facilities. Each of the following actionable conclusions is intended to lead to positive outcomes taken alone, but the full implementation of all actionable conclusions taken together amplifies the impact as the set is interrelated, self-consistent, complementary, and mutually supporting.

\subsection{UF Conclusion \#1: Excavate now for future experiments}
\label{sec:uf-recommendation-1}
Evaluations from the Snowmass Underground Facilities and Infrastructure Frontier's efforts drew a conclusion that today's underground facilities are currently full to capacity with current and planned experiments. More underground space is required for the research planned for the late 2020s and beyond. The mobilization cost for underground excavation is on the order of a small-to-mid-scale experiment (tens of million USD). Acting now to extend excavation at SURF can cost-effectively provide the necessary space to fulfill the science objectives of the Neutrino and Cosmic Frontiers in the coming decade and beyond.
\begin{quote}
    \textbf{UF Conclusion \#1: Leverage the Long Baseline Neutrino Facility excavation enterprise to increase underground space at Sanford Underground Research Facility in a timely and cost-effective way to permit siting of next-generation underground high energy physics research experiments.}
\end{quote}
The high energy physics community is planning to pursue both large- and small-to-mid-scale underground searches for the direct detection of particle-like dark matter. A large-scale liquid noble dark matter experiment will require a host cavern on the 25$\times$25$\times$25~m$^{3}$ size, ready and available on the time scale of the late 2020s. Small-to-mid-scale dark matter experiments, for instance those part of the U.S. DOE Office of High Energy Physics' Dark Matter New Initiative efforts, would benefit from outfitted cavern drifts able to support multiple experiments, ready and available on the time scale of the mid-2020s. The excavations required for these effort are comparable to a single module of the Long Baseline Neutrino Facility.

Connecting to UF Conclusion \#2, the designation of SURF as a U.S. DOE User Facility is not fully realized and does not provide utility to the broader scientific community \emph{unless} space for research is ready, outfitted, and organized under an appropriate managing institutional structure.

\subsection{UF Conclusion \#2: U.S. Underground User Facility}
\label{sec:uf-recommendation-2}
The Sanford Underground Research Facility (SURF) is both the single largest underground science-oriented facility in the U.S. supporting high energy physics and the location of the current largest U.S. investment in infrastructure for long-baseline neutrino detection. Furthermore, SURF hosts dark matter and neutrinoless double-beta decay experiments, both seeking to probe physics beyond the Standard Model. Finally, SURF has provided space and support for a range of other research from accelerator-based nuclear astrophysics to geoengineering. Future users may include quantum information science (QIS) researchers, atom interferometry gravitational wave detection, and/or biological science.
\begin{quote}
    \textbf{UF Conclusion \#2: Designate the Sanford Underground Research Facility as a U.S. Department of Energy User Facility.}
\end{quote}
The U.S. Department of Energy's Office of Science manages a formal process for the designation of a user facility: ``A user facility is a federally sponsored research facility available for external use to advance scientific or technical knowledge\ldots'' meeting a number of conditions. Designation of SURF as a user facility would provide the high energy physics community a clear structure for seeking hosted underground space and infrastructure use agreements that will accelerate the research community's ability to plan and conduct next-generation experiments requiring underground locations. A list of existing \href{https://www.energy.gov/science/office-science-user-facilities}{DOE Office of Science user facilities}~\cite{UserFacilities} is available as well as the \href{https://science.osti.gov/User-Facilities/Policies-and-Processes/Definition}{process and definition}~\cite{UserFacilityProcess} for creation of such user facilities.

\subsection{UF Conclusion \#3: Support completion of LBNF and DUNE}
\label{sec:uf-recommendation-3}
As described by Lia Merminga, Director of Fermi National Accelerator Laboratory, the U.S. will soon host the world's most capable neutrino experiment composed of the Long Baseline Neutrino Facility (LBNF) and the Deep Underground Neutrino Experiment (DUNE). To ensure success of this flagship U.S. high energy physics program in long-baseline neutrino research for the coming two decades, the following conclusion is made:
\begin{quote}
    \textbf{UF Conclusion \#3: Provide full support for the underground facilities hosting the Long Baseline Neutrino Facility (LBNF) and the Deep Underground Neutrino Experiment (DUNE).}
\end{quote}
The program of research pursued by LBNF and DUNE is central to the Snowmass Neutrino Frontier's vision for scientific discovery in the coming two decades. Organizational and infrastructure support for the hosting underground facility during this time period is a key component to ensuring this flagship U.S. high energy physics research program delivers. Achieving the full scientific reach of this long-baseline neutrino research program will require completion of all four planned DUNE modules, an investment in discovery science reaching into the 2030s time period.

Additionally, connecting to UF Conclusions \#1 and \#2, the excavation of space and designation of SURF as a U.S. DOE User Facility, respectively, provides an integrated means for evaluating, managing, and overseeing any proposed beneficial utilization of the LBNF caverns in advance of the full and complete installation of DUNE modules 3 and 4 in the 2030s.

\subsection{UF Conclusion \#4: Realize the 2014 P5 Recommendation for G3 dark matter}
\label{sec:uf-recommendation-4}
In 2021--2022, the Snowmass Cosmic Frontier continues to place high priority on the search for dark matter. In this respect, the 2014 Particle Physics Project Prioritization Panel (P5) Recommendation 20 is yet unfulfilled. The full supporting text of the 2014 P5 Recommendation 20 is:
\begin{quote} \begin{quote}
\textit{``The results of G2 direct detection experiments and other contemporaneous dark matter searches will guide the technology and design of third-generation experiments. As the scale of these experiments grows to increase sensitivity, the experimental challenge of direct detection will still require complementary experimental techniques, and international cooperation will be warranted. The U.S. should host at least one of the third-generation experiments in this complementary global suite.}

\textit{\textbf{Recommendation 20: Support one or more third-generation (G3) direct detection experiments, guided by the results of the preceding searches. Seek a globally complementary program and increased international partnership in G3 experiments.}''}
\end{quote} \end{quote}
Considering the Cosmic Frontier’s continued priorities requiring underground facilities, and the 2014 P5 Recommendation 20, the following conclusion is made:
\begin{quote}
    \textbf{UF Conclusion \#4: Following the 2014 P5 Recommendation 20, R\&D and decision making for a third-generation direct detection dark matter program should commence immediately to enable a construction start in the late 2020s.}
\end{quote}
This conclusion drives the planning and decision making process that will determine the underground facility and infrastructure requirements to support experiments of these scales. As the third-generation particle-like dark matter program is projecting construction start in the late 2020s, the preceding development of underground facilities and infrastructure must begin now.

UF Conclusion \#4 is highly synergistic with UF Conclusion \#1, but UF Conclusion \#4 must be pursued of its own accord to ensure the search for particle-like dark matter, one of the Cosmic Frontier's highest priorities is ready to progress in the late 2020s and into the 2030s time period.

\subsection{UF Conclusion \#5: Support underground researchers and R\&D}
\label{sec:uf-recommendation-5}
The nature and maturity of high energy physics research performed in underground facilities is reaching a point of becoming \emph{generational}, where experiments are planed, constructed, and performed over multiple-decade time scales. A diverse and inclusive workforce educated, skilled, and active in the development of the tools and techniques used in underground high energy physics experiments is necessary to carry these experiments through to completion.
\begin{quote}
    \textbf{UF Conclusion \#5: To ensure a robust collection of scientific programs in underground facilities, support the enabling capabilities, technique development, and expertise required for underground experiments.}
\end{quote}
This conclusion is focused on ensuring research funding support is available to train the next-generation of underground researchers through experiential mentoring from today's underground scientific experts. In addition to the construction of major projects, small-scale R\&D programs to support advances in techniques and capabilities supporting underground science are a critical component of maintaining the field and educating the next generation.

UF Conclusion \#5 undergirds the other UF Conclusions \#1 through \#4, for underground facilities and infrastructure are nothing without the researchers who invest their effort and careers in the discovery science that is only possible when performed below the Earth's surface.








\bibliographystyle{JHEP}
\bibliography{myreferences} 

\end{document}